\documentclass[aps,twocolumn,showpacs,prl,superscriptaddress]{revtex4}
\usepackage{epsf}
\usepackage{color}

\begin{document}
\title{Microcanonical entropy inflection points: Key to systematic understanding\\ 
of transitions in finite systems}
\author{Stefan Schnabel}
\email[E-mail: ]{stefanschnabel@physast.uga.edu}
\affiliation{Center for Simulational Physics, The University of Georgia, Athens,
Georgia 30602, USA}
\author{Daniel T.\ Seaton}
%\email[E-mail: ]{dseaton@hal.physast.uga.edu}
\email[E-mail: ]{dseaton@mit.edu}
\affiliation{Center for Simulational Physics, The University of Georgia, Athens,
Georgia 30602, USA}
\affiliation{Department of Physics, Massachusetts Institute of Technology,
Cambridge, Massachusetts 02139, USA}
\author{David P.\ Landau}
\email[E-mail: ]{dlandau@hal.physast.uga.edu}
\affiliation{Center for Simulational Physics, The University of Georgia, Athens,
Georgia 30602, USA}
\author{Michael Bachmann}
\email[E-mail: ]{bachmann@smsyslab.org}
\homepage[\\ Homepage: ]{http://www.smsyslab.org}
\affiliation{Center for Simulational Physics, The University of Georgia, Athens,
Georgia 30602, USA}
\begin{abstract}
We introduce a systematic classification method for the analogs of phase transitions
in finite systems. This completely general analysis, which is applicable to any
physical system and extends towards the thermodynamic limit,
is based on the microcanonical entropy and its energetic derivative, the inverse caloric
temperature. Inflection points of this quantity signal 
cooperative activity and thus serve as distinct indicators of transitions.
We demonstrate the power of this method through application to the long-standing problem
of liquid-solid transitions in elastic, flexible homopolymers.
\end{abstract}
\maketitle
Structure formation processes are typically accompanied by nucleation transitions, where
crystalline shapes form out of a liquid or vapor phase. Thus, nucleation is
governed by finite-size and surface effects. For small physical systems,
it is difficult to understand thermodynamic transitions of this type, as they
strongly depend on system size.

Cooperativity refers to collective
changes in a statistically significant fraction of the degrees of freedom in a system, which
transforms the system into a new macrostate. In the thermodynamic limit of an
infinitely large system, the ensemble of macrostates sharing similar
thermodynamic properties would be called a ``phase'' and the transformation a
``phase transition''. The description of such a transformation in a
\emph{finite} system is more subtle, as it cannot be described in the
traditional Ehrenfest scheme of singularities in response quantities. However,
statistical physics and thus thermodynamics are also valid for systems 
with no thermodynamic limit. Examples include the structure
formation in small atomic clusters and all biomolecules. This is particularly
striking for proteins, i.e., heterogeneous linear chains of amino acids. The
fact that the individual biological function is connected with the geometrical
shape of the molecule makes it necessary to discriminate unfolded (non-functional)
and folded (functional) states. Although these systems are finite, they undergo a
structural transition by passing a single (or more) free-energy
barrier(s). Since these finite-system transitions exhibit strong similarities
compared to phase transitions, we extend the terminology once defined in the
thermodynamic limit to all systems exhibiting cooperative behavior.

In this paper, we introduce a commonly applicable and simple method for the
identification and classification of cooperative behavior in systems of arbitrary size by
means of microcanonical thermodynamics~\cite{gross1}. 
It also includes the precise and straightforward analysis of
the finite-size effects,
which are important to a general understanding of the onset of phase transitions.
This is in contrast to canonical
approaches, where detailed information is lost by
averaging out thermal fluctuations. Re-gaining information
about finite-size effects in canonical schemes, e.g., by the investigation 
of the distribution of
Lee-Yang zeros in the complex temperature plane~\cite{borrmann} or by inverse
Laplace transform~\cite{cohen1}, is complicated.

The identification of transitions is associated with a distinct
definition of transition points such as a transition temperature. In the
canonical representation of finite systems these usually differ, e.g., peak structures
of thermodynamic quantities, such as the specific heat and fluctuations of order parameters 
as functions of the heat-bath temperature. 
This makes it impossible to fix a unique transition
point. In the microcanonical analysis, the temperature is defined via the curvature of the caloric
entropy curve and thus all transition signals in the microcanonical entropy can be
directly associated with a transition temperature.

After introducing the method, we apply it to liquid-solid and solid-solid transitions occurring
for elastic, flexible polymers, which have been under debate for quite some time.
In contrast to the rather well-understood coil-globule
collapse transition, the formation of highly compact crystalline, amorphous, or glasslike
structures intricately depends on the precise relation of intrinsic 
energy and length scales in the system~\cite{binder0,vbj1,seaton1,sbj1,taylor1,seaton3}.

In recent work, the microcanonical analysis has successfully
been applied in aggregation studies of coarse-grained polymer and peptide
models, where a nucleation process
was found to be an energetically ordered hierarchy of individual structural subphase
transitions~\cite{jbj1}. Caloric approaches have also been used to investigate
the folding behavior of proteins~\cite{liang1,gomez1,b1,bbd1,bdb1}
and the structural phases of polymers with stiff bonds~\cite{taylor1},
as well as polymer adsorption transitions~\cite{bj1,liang3,mjb1}.
Other applications include the formation of galaxies~\cite{thirring1},
the clustering and fragmentation
of atomic clusters and nuclei~\cite{gross1,gross2,doye1}, and
order--disorder transitions in spin systems~\cite{gross1,wj1,hueller1,cohen1,pleimling1}.
Most of these studies are aimed at using the microcanonical analysis as an alternative approach
to investigating finite-size scaling properties. However, a systematic scheme for the
classification of transitions in the respective 
\emph{finite systems} has remained lacking. The method introduced here
closes this gap through the introduction of an
Ehrenfest-like analysis based on microcanonical entropy inflection points.

A fundamental property of each physical system, and the central quantity for 
our method, is the microcanonical entropy $S(E)=k_B\ln g(E)$, where
$g(E)$ is the density of states for a given energy $E$ 
(in the following, we will set $k_B\equiv 1$). 
Alternatively, a volume entropy can be defined via the integrated density
of states
by $S'(E)=k_B\ln G(E)$ with
$G(E)=\int_{E_{\rm min}}^EdE' g(E')$~\cite{hertz1}, which is virtually identical with $g(E)\Delta E$
($|\Delta E/E|\ll 1$) in the transition regions~\cite{jbj1}. It has been argued
that only $G(E)$ is consistent with
the classical equipartition
theorem~\cite{tiller1,campisi1}, however,
its physical meaning is much less obvious~\cite{gross1}. Therefore, we will
continue using $S(E)$ instead. It should also be mentioned that $g(E)$ is
the ``natural'' output provided, e.g., by generalized-ensemble Monte Carlo methods.
Among the most prominent of these methods are
multicanonical~\cite{muca} and Wang-Landau~\cite{wl}
sampling, which enable a precise numerical estimation of this quantity 
over hundreds or even thousands of orders of
magnitude~\cite{sbj1,seaton1}.

A qualitative change in the interplay of entropy and energy
in the system is signaled by noticeable alterations in the curvature of $S(E)$, 
which are quantitative measures for the strength of cooperativity
of the associated transitions.
For finite systems exhibiting transitions with phase separation, $S(E)$ can even
possess convex regions~\cite{gross1}, although it is a strictly concave function
in the thermodynamic limit. In this case, 
the slope of a tangent at each point of the curve is unique, and it is common
to define the reciprocal microcanonical temperature via the caloric derivative of $S$,
\begin{equation}
\label{eq:beta}
\beta(E)\equiv T^{-1}(E)=(dS/dE)_{N,V}~, 
\end{equation}
where system size $N$ and volume $V$ are kept constant.
In the thermodynamic limit, where fluctuations about the mean energy 
become negligible, the canonical and microcanonical ensembles are identical,
and the canonical (or heat-bath) temperature equals the microcanonical temperature. 
This is not the 
case for a finite system experiencing a structural transition, where different
quantities vary in their fluctuation properties, rendering an identification of transition points impossible. 
Since the complete phase behavior is already encoded in $S(E)$, 
it is useful to consider $\beta(E)$ as a unique parameter to identify transition points.

We further propose to analyze the monotonic behavior of $\beta(E)$,
expressed by its derivative with respect to energy,
\begin{equation}
\label{eq:gamma}
\gamma(E)=d\beta(E)/dE = d^2S/dE^2. 
\end{equation}
This will allow for the introduction of a systematic classification scheme 
of transitions in finite systems.
In principle, this can also be used for scaling analyses towards the thermodynamic limit.
\begin{figure}
\centerline{\epsfxsize=8.8cm \epsfbox{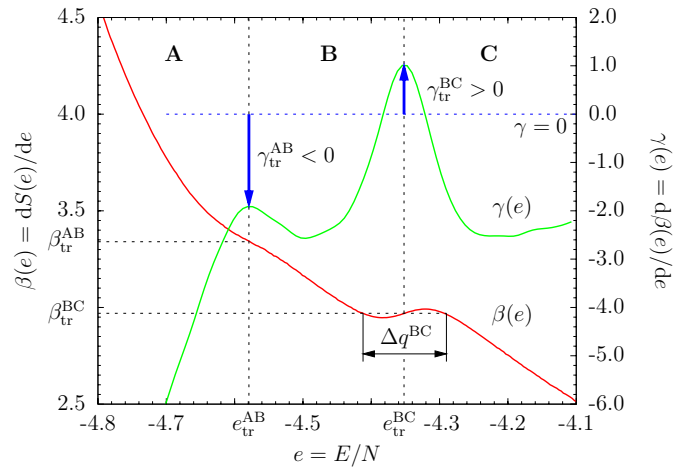}}
\caption{\label{fig:infl} 
(Color online) Inverse temperature $\beta(e)$ and its derivative $\gamma(e)$ as functions
of the energy per particle, $e=E/N$, exemplified for an elastic polymer with 102 monomers. The maxima of $\gamma(e)$ indicate transitions between
the structural phases A and B at $e_{\rm tr}^{\rm AB}$ and B and C at $e_{\rm tr}^{\rm BC}$. The associated points $\beta(e_{\rm tr}^{\rm AB})= \beta_{\rm tr}^{\rm AB}$
and $\beta(e_{\rm tr}^{\rm BC})= \beta_{\rm tr}^{\rm BC}$ define the transition temperatures
$T_{\rm tr}^{\rm AB}=(\beta_{\rm tr}^{\rm AB})^{-1}$ and 
$T_{\rm tr}^{\rm BC}=(\beta_{\rm tr}^{\rm BC})^{-1}$. According to our 
classification scheme, the transition between A and B is of second order, since the 
slope of the inflection point is negative. On the other hand, B$\leftrightarrow$C is a first-order
transition as the respective slope at $\beta(e_{\rm tr}^{\rm BC})$ is positive. The non-monotonicity
of $\beta(e)$ in this region, called ``backbending'', is a typical signal of phase coexistence.
The latent heat $\Delta q^{\rm BC}$ is defined as the energetic width of this transition region.
}
\vspace*{-5mm}
\end{figure}

We define a transition between phases to be of \emph{first order} if the slope of the 
corresponding inflection point of $\beta(E)$ at $E=E_{\rm tr}$ is positive,
i.e., $\gamma_{\rm tr}=\gamma(E_{\rm tr}) > 0$. Only in this case is the 
temperature curve non-monotonic and there is no unique mapping between
$\beta$ and $E$. Physically, both phases coexist in the transition region.
The overall energetic width of 
the undercooling, backbending, and overheating regions, obtained from a Maxwell construction,
is thus identical to the latent heat. Therefore, for a first-order transition, $\Delta q > 0$.
In the case that the inflection point has a negative slope, $\gamma_{\rm tr}=\gamma(E_{\rm tr}) < 0$,
the phases cannot coexist and the latent heat is zero. 
In complete analogy to phase transitions in the 
thermodynamic limit, we classify such transitions as of \emph{second order}.
Since the inflection points of $\beta(E)$ correspond to maxima
in $\gamma(E)$, it is therefore sufficient to analyze the peak structure of $\gamma(E)$
in order to identify the transition energies and temperatures. The sign of the peak values 
classifies the transition. This very simple and general classification scheme
applies to all physical systems. 

Figure~\ref{fig:infl} illustrates the procedure for the identification of the
transitions by means of inflection-point analysis, where the inverse temperature
$\beta$ and its energetic derivative $\gamma$ are plotted as functions of the reduced energy
$e=E/N$, with $N$ being the system size.
As a first example, we consider an elastic flexible homopolymer
with $N=102$ monomers. This system
exhibits four structural phases~\cite{sbj1}: two solid icosahedral phases (A: Mackay, B: anti-Mackay), 
a globular liquid phase (C), and the random-coil phase (D).  In Fig.~\ref{fig:infl},
the transitions can indeed be uniquely
identified (since the C$\leftrightarrow$D transition occurs at much higher energy and 
temperature, it is not included, but can also easily be found
by inflection-point analysis; it is a second-order transition at 
$e_{\rm tr}^{\rm CD}\approx -1.21$, $\beta_{\rm tr}^{\rm CD}\approx 1.08$).
A 1$^{\rm st}$-order liquid-solid transition B$\leftrightarrow$C is characterized by $\gamma_{\rm tr}^{\rm BC}>0$
at $e_{\rm tr}^{\rm BC}\approx -4.35$ ($\beta_{\rm tr}^{\rm BC}\approx 2.97$). 
The width of the energetic 
transition region 
corresponds to the latent heat, $\Delta q^{\rm BC}$, which is obviously nonzero because of 
the backbending effect or the coexistence of both phases in this region. The
2$^{\rm nd}$-order 
transition A$\leftrightarrow$B is found at $e_{\rm tr}^{\rm AB}\approx -4.58$ 
($\beta_{\rm tr}^{\rm AB}\approx 3.34$) 
by an inflection point with negative slope ($\gamma_{\rm tr}^{\rm AB}< 0$).
\begin{figure}
\centerline{\epsfxsize=8.8cm \epsfbox{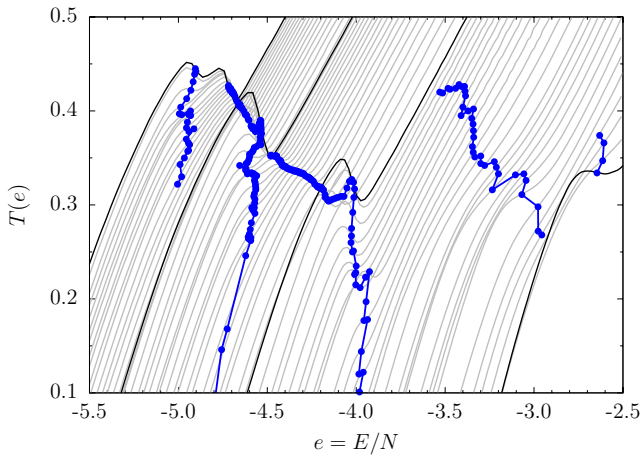}}
\caption{\label{fig:tcal} 
(Color online)
Caloric temperature curves $T(e)=\beta^{-1}(e)$ for a selection of elastic, flexible polymers 
with chain lengths in the interval $N=13,\ldots,309$ (from right 
to left). Curves for chains with magic length ($N=13,55,147,309$) are bold. 
The relevant inflection points, indicating the conformational transitions
on the basis of our analysis method, are marked by \textcolor{blue}{$\bullet$} symbols.
}
\vspace*{-5mm}
\end{figure}

In order to demonstrate the capability of our method to systematically analyze all
transitions in finite systems, we estimate the transition points for the entire set of elastic 
Lennard-Jones homopolymers with 
$N=13,\ldots,309$ monomers. In the liquid and solid regimes, the structural behavior 
of these polymers is very similar to rare-gas systems consisting of $N$ atoms, which also form
compact, crystalline clusters at very low temperatures~\cite{frant1,sbj1,seaton3}.
We employ the standard model for flexible, elastic polymers, where
the monomers interact via a truncated-shifted 
Lennard-Jones potential, $E_{\rm LJ}^{\rm mod}(r_{ij})=E_{\rm LJ}(\min(r_{ij},r_{\rm c}))-
E_{\rm LJ}(r_{\rm c})$
with $E_{\rm LJ}(r_{ij})=4\epsilon[(\sigma/r_{ij})^{12}-(\sigma/r_{ij})^{6}]$, where
$r_{ij}$ is the distance between two monomers located at ${\bf r}_i$ and ${\bf r}_j$ 
($i,j=1,\ldots,N$),
and
$\epsilon=1$ and $\sigma=2^{-1/6}r_0$, with the potential minimum at $r_0=0.7$ and the cutoff at 
$r_{\rm c}=2.5\sigma$.
Adjacent monomers are connected by finitely extensible nonlinear 
elastic (FENE) anharmonic bonds~\cite{FENE,binder3}, 
$E_{\rm FENE}(r_{i\,i+1})=-KR^2\ln\{1-[(r_{i\,i+1}-r_0)/R]^2\}^{1/2}$.  
The FENE potential minimum is located at $r_0$ and diverges for 
$r\rightarrow r_0\pm R$ (in our simulations $R=0.3$). The spring constant $K$ is set to $40$.
The total energy of a polymer conformation ${\bf X}=({\bf r}_1,\ldots,{\bf r}_N)$ is
given by
$E({\bf X})=\sum_{i=1}^N\sum_{j=i+1}^NE^{\rm mod}_{\rm LJ}(r_{ij})+\sum_{i=1}^{N-1}E_{\rm FENE}(r_{ii+1}).$
\begin{figure}
\centerline{\epsfxsize=8.8cm \epsfbox{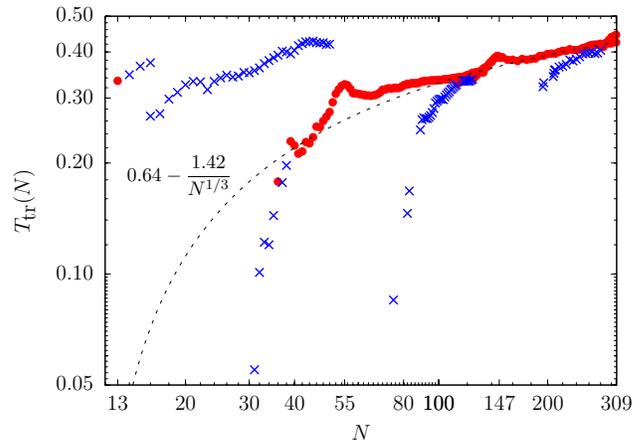}}
\caption{\label{fig:tn} 
(Color online)
Transition temperatures $T_{\rm tr}(N)$ of conformational transitions for small elastic polymers 
with chain lengths $N=13,\ldots,309$ in the liquid-solid and solid-solid transition
regimes, obtained from inflection-point analysis. 
First-order transition points are marked by red symbols (\textcolor{red}{$\bullet$}), 
second-order transition points by blue symbols (\textcolor{blue}{$\times$}).
Also shown is a fit for the liquid-solid transition temperature towards the 
thermodynamic limit $N\to \infty$ (dashed line).
}
\vspace*{-5mm}
\end{figure}

Figure~\ref{fig:tcal} shows the
caloric temperature curves for elastic polymers with various chain lengths
in the liquid and solid regimes, calculated from highly accurate density of states estimates
obtained in sophisticated multicanonical Monte Carlo
simulations~\cite{sjb1}. The identified inflection points associated with conformational
transitions are indicated by \textcolor{blue}{$\bullet$} symbols. 
As expected, there is no general and 
obvious relation of the behavior
of chains with slightly different lengths. This is due to the still dominant finite-size 
effects of the polymer trying to reduce their individual surface-to-volume
ratio,
which therefore strongly depends on optimal monomer packings in the interior and on the surface 
of the conformations. For example, for chains of moderate lengths 
($N\le 147$~\cite{sbj1,seaton3}), 
the different behavior can be
traced back to the monomer arrangements on the facets of icosahedral
structures, known as Mackay and anti-Mackay overlayers~\cite{northby1}. Solid-solid 
transitions between Mackay and anti-Mackay structures are also possible under
certain conditions in these systems~\cite{sbj1,seaton3}. This can be seen
in Fig.~\ref{fig:tn}, where all transition temperatures 
$T_{\rm tr}(N)=\beta^{-1}_{\rm tr}(N)$ for liquid-solid and solid-solid
transitions are plotted in dependence of the chain length $N$~\cite{rem1}. 
Symbols \textcolor{red}{$\bullet$} indicate first-order transitions, which for $N>38$ can be associated to
the respective liquid-solid transitions, whereas 
symbols \textcolor{blue}{$\times$} mark second-order transitions.

If the associated transition temperatures
are smaller than the liquid-solid transition temperatures, the symbols indicating 
second-order behavior belong to solid-solid transitions, e.g., 
transitions between geometrical shapes with Mackay or anti-Mackay overlayers.
Note the different behavior for ``magic'' chain lengths 
$N_{\rm magic}=13,55,147,309,\ldots$,
in which icosahedral Mackay ground states form.
Figure~\ref{fig:tn} also gives evidence for the convergence of the solid-solid 
and liquid-solid transition temperatures when $N$ approaches 
a magic length. This behavior repeats for each $N$ interval that finally ends 
at a certain magic length $N_{\rm magic}$, where both transitions merge into a 
single first-order liquid-solid transition. 
The influence of the solid-solid effects 
weakens with increasing system size, while the liquid-solid transition remains a true 
phase transition in the thermodynamic limit.
Inserted into the plot is a fit function $T_{\rm tr}(N)=T_{\rm}^{\rm ls}-aN^{-1/3}$
which suggests an estimate for the thermodynamic phase transition temperature 
$T_{\rm tr}^{\rm ls}\approx 0.64$.

Summarizing, we have introduced a general method for the analysis of phase 
transitions in small systems based on the central quantity of any statistical system, the
microcanonical entropy, and applied it to the long-standing problem of
structural transitions of flexible polymers. Advanced Monte Carlo simulation techniques such as multicanonical
sampling~\cite{muca} and the Wang-Landau method~\cite{wl} enable precise estimations of
the density of states, and thus it is straightforward to obtain the microcanonical entropy
in computer simulations.
Since indicative quantities such as transition temperatures can be
quantitatively determined, our method also enables experimentally competitive 
predictions.
%

%\acknowledgments
This project has been partially supported by NSF DMR-0810223.

\begin{thebibliography}{199}
%
\bibitem{gross1}
D.\ H.\ E.\ Gross, \emph{Microcanonical Thermodynamics} (World Scientific, Singapore, 2001).
%
\bibitem{borrmann}
P.\ Borrmann, O.\ M\"ulken, and J.\ Harting, Phys.\ Rev.\ Lett.\ \textbf{84},
3511 (2000); O.\ M\"ulken, H.\ Stamerjohanns, and P.\ Borrmann, Phys.\ Rev.\ E
\textbf{64}, 047105 (2001).
%
\bibitem{cohen1}
I.\ Ispolatov and E.\ G.\ D.\ Cohen, Physica A \textbf{295}, 475 (2001).
%
\bibitem{binder0}
F.\ Rampf, W.\ Paul, K.\ Binder, Europhys.\ Lett.\ \textbf{70}, 628 (2005);
J.\ Polym.\ Sci.: Part B: Polym.\ Phys.\ \textbf{44}, 2542 (2006);
W.\ Paul, T.\ Strauch, F.\ Rampf, and K.\ Binder, Phys.\ Rev.\ E \textbf{75}, 060801(R) (2007).
%
\bibitem{vbj1}
T.\ Vogel, M.\ Bachmann, and W.\ Janke, Phys.\ Rev.\ E \textbf{76}, 061803 (2007).
%
\bibitem{seaton1}
D.\ T.\ Seaton, S.\ J.\ Mitchell, and D.\ P.\ Landau, Braz.\ J.\ Phys.\
\textbf{38}, 48 (2008); D.\ T.\ Seaton, T.\ W\"ust, and D.\ P.\ Landau, Comp.\ Phys.\ Comm.\ \textbf{180}, 587 (2009).
%
\bibitem{sbj1}
S.\ Schnabel, T.\ Vogel, M.\ Bachmann, and W.\ Janke, Chem.\ Phys.\ Lett.\
\textbf{476}, 201 (2009); S.\ Schnabel, M.\ Bachmann, and W.\ Janke, J.\ Chem.\
Phys.\ \textbf{131}, 124904 (2009).  
%
\bibitem{taylor1}
M.\ P.\ Taylor, W.\ Paul, and K.\ Binder, Phys.\ Rev.\ E \textbf{79}, 050801(R) (2009);
J.\ Chem.\ Phys.\ \textbf{131}, 114907 (2009).
%
\bibitem{seaton3}
D.\ T.\ Seaton, T.\ W\"ust, and D.\ P.\ Landau, Phys.\ Rev.\ E \textbf{81}, 011802 (2010).
%
\bibitem{jbj1}
C.\ Junghans, M.\ Bachmann, and W. Janke, Phys.\ Rev.\ Lett.\ \textbf{97}, 218103 (2006);
J.\ Chem.\ Phys.\ \textbf{128}, 085103 (2008);
Europhys.\ Lett.\ \textbf{87}, 40002 (2009).
%
\bibitem{liang1}
T.\ Chen, X.\ S.\ Lin, Y.\ Liu, and H.\ J.\ Liang, Phys.\ Rev.\ E \textbf{76}, 046110 (2007);
Phys.\ Rev.\ E \textbf{78}, 056101 (2008).
%
\bibitem{gomez1}
J.\ Hern\'andez-Rojas and J.\ M.\ Gomez-Llorente, Phys.\ Rev.\ Lett.\ \textbf{100}, 258104 (2008).
%
\bibitem{b1}
M.\ Bachmann, Phys.\ Proc.\ \textbf{3}, 1387 (2010).
%
\bibitem{bbd1}
T.\ Bereau, M.\ Bachmann, and M.\ Deserno, J.\ Am.\ Chem.\ Soc.\ \textbf{132},
13129 (2010).
%
\bibitem{bdb1}
T.\ Bereau, M.\ Deserno, and M.\ Bachmann, Biophys.\ J.\, in press (2011).
%
\bibitem{bj1}
M.\ Bachmann and W.\ Janke, Phys.\ Rev.\ Lett.\ \textbf{95}, 058102 (2005);
Phys.\ Rev.\ E \textbf{73}, 041802 (2006); Lect.\ Notes Phys.\ \textbf{736}, 203 (2008).
%
\bibitem{liang3}
L.\ Wang, T.\ Chen, X.\ S.\ Lin, Y.\ Liu, and H.\ J.\ Liang, J.\ Chem.\ Phys.\ \textbf{131}, 244902 (2009).
%
\bibitem{mjb1}
M.\ M\"oddel, W.\ Janke, and M.\ Bachmann, Phys.\ Chem.\ Chem.\ Phys., \textbf{12}, 11548 (2010).
%
\bibitem{thirring1}
W.\ Thirring, Z.\ Physik \textbf{235}, 339 (1970).
%
\bibitem{gross2}
D.\ H.\ E.\ Gross and J.\ F.\ Kenney, J.\ Chem.\ Phys.\ \textbf{122}, 224111 (2005).
%
\bibitem{doye1}
E.\ G.\ Noya and J.\ P.\ K.\ Doye, J.\ Chem.\ Phys.\ \textbf{124}, 104503 (2006).
%
\bibitem{wj1}
W.\ Janke, Nucl.\ Phys.\ B (Proc.\ Suppl.) \textbf{63A-C}, 631 (1998).
%
\bibitem{hueller1}
M.\ Kastner, M.\ Promberger, and A.\ H\"uller, J.\ Stat.\ Mech.\ \textbf{99}, 1251 (2000).
%
\bibitem{pleimling1}
H.\ Behringer and M.\ Pleimling, Phys.\ Rev.\ E \textbf{74}, 011108 (2006);
H.\ Behringer, Entropy \textbf{10}, 224 (2008).
%
\bibitem{hertz1}
P.\ Hertz, Ann.\ Phys.\ \textbf{33}, 225 (1910); \textit{ibid.}, 537 (1910).
%
\bibitem{tiller1}
E.\ M.\ Pearson, T.\ Halicioglu, and W.\ A.\ Tiller, Phys.\ Rev.\ A \textbf{32},
3030 (1985).
%
\bibitem{campisi1}
M.\ Campisi and D.\ H.\ Kobe, Am.\ J.\ Phys.\ \textbf{78}, 608 (2010).
%
\bibitem{muca}
B.\ A.\ Berg and T.\ Neuhaus, Phys.\ Lett.\ B \textbf{267}, 249 (1991);
Phys.\ Rev.\ Lett.\ \textbf{68}, 9 (1992).
%
\bibitem{wl}
F.\ Wang and D.\ P.\ Landau, Phys.\ Rev.\ Lett.\ \textbf{86}, 2050 (2001);
Phys.\ Rev.\ E \textbf{64}, 056101 (2001);
Comp.\ Phys.\ Commun.\ \textbf{147}, 674 (2002).
%
\bibitem{frant1}
P.\ A.\ Frantsuzov and V.\ A.\ Mandelshtam, Phys.\ Rev.\ E\  \textbf{72}, 037102 (2005).
%
\bibitem{FENE}
R.\ B.\ Bird, C.\ F.\ Curtiss, R.\ C.\ Armstrong, and O.\ Hassager, 
\emph{Dynamics of Polymeric Liquids}, 2nd ed., 2 vols.\ (Wiley, New York, 1987).
%
\bibitem{binder3}
A.\ Milchev, A.\ Bhattacharaya, and K.\ Binder, Macromolecules \textbf{34}, 1881 (2001).
%
\bibitem{sjb1}
S.\ Schnabel, W.\ Janke, and M.\ Bachmann, J.\ Comput.\ Phys.\ \textbf{230},
4454 (2011).
%
\bibitem{northby1}
J.\ A.\ Northby, J.\ Chem.\ Phys.\ \textbf{87}, 6166 (1987).
%
\bibitem{rem1}
The inflection-point analysis for $\beta(E)$ can also easily be applied
to the coil-globule  
transition, which occurs at much higher temperatures (for this reason, transition points are not inserted in Fig.~\ref{fig:tn}).
%
\end{thebibliography}
\end{document}